\documentclass[twocolumn,prl]{revtex4}

\ifx\pdfoutput\undefined
  \usepackage[dvips]{graphicx}
\else
  \usepackage[pdftex]{graphicx}
\fi

\usepackage{dcolumn}
\usepackage{amsmath}
\usepackage{latexsym}

\begin{document}

\title[Short Title]{Bloch inductance in small-capacitance Josephson junctions}
\author{A.~B.~Zorin}
\affiliation{Physikalisch-Technische Bundesanstalt, Bundesallee
100, 38116 Braunschweig, Germany}%

\date{October 17, 2005}

\begin{abstract}
We show that the electrical impedance of a small-capacitance Josephson junction
includes besides the capacitive term $-i/\omega C_B$ also an inductive term
$i\omega L_B$. Similar to the known Bloch capacitance $C_B(q)$, the Bloch
inductance $L_B(q)$ also depends periodically on the quasicharge $q$, and its
maximum value achieved at $q=e \,(\textrm{mod}\:2e)$ always exceeds the value
of the Josephson inductance of this junction $L_{J}(\varphi)$ at fixed
$\varphi=0$. The effect of the Bloch inductance on the dynamics of a single
junction and a one-dimensional array is described.

\verb  PACS numbers: 74.50.+r, 03.75.Lm, 74.81.Fa
\end{abstract}
\maketitle

In the recent years the Josephson tunnel junctions with small capacitance have
been extensively studied. The interest in these junctions and in circuits with
these junctions can be explained by the remarkable quantum behavior of these
macroscopic systems. For example, the small-capacitance junctions can transfer
individual Cooper pairs and generate Bloch oscillations
\cite{LikZor,AZL,KuzHav}. These junctions can also exhibit a long-time quantum
coherence \cite{Nakam,Vion}. Important potential applications of these effects
include the fundamental standard of current \cite{LZ-IEEE} and quantum
computing circuits \cite{Makhlin}.

The small-capacitance Josephson junctions are characterized by the finite ratio
$\lambda=E_J/E_c$  of the Josephson coupling energy $E_J = (\Phi_0/2\pi)I_c$
(here $I_c$ is the critical current and $\Phi_0=h/2e$ is the flux quantum) and
the charging energy $E_c=e^2/2C$ ($C$ is the junction capacitance). The
remarkable feature of these junctions is their nonlinear differential
capacitance $C_B$ related to the local curvature of the zero Bloch energy band
\cite{LikZor,AZL},
\begin{equation}
\label{C_B} C_B^{-1}(q) = \frac{d^2 E_0}{dq^2},
\end{equation}
where $E_0(q)$ is the ground state energy of the junction. $E_0$ periodically
(the period is equal to $2e$) depends on the quasicharge $q = \int^{t}I(t')
dt'$, which is a good variable driven by the current source $I$ (see Fig.\,1a).
Variable $q$ is analog to the quasimomentum of electron moving in the periodic
potential of a crystal lattice. Similar to the Josephson inductance
$L_J(\varphi)$, which depends on the classical phase $\varphi$  in the large
junctions ($\lambda \rightarrow \infty$), capacitance $C_B$ can also take on
both positive and negative values. The property of variation of the Bloch
capacitance in a large range was turned to advantage by Averin and Bruder
\cite{AvBrud} who suggested a switchable capacitive coupling between Josephson
charge qubits. Incorporating the nonlinear Bloch capacitance into a tank
circuit made it possible to realize a sensitive electrometer \cite{Roschier}
and efficient readout of the Cooper-pair-box qubit \cite{Duty}.

Behavior of these circuits is usually described by the junction model
comprising solely the Bloch capacitance Eq.\,(\ref{C_B}). Actually, this
approach is applicable for the static case or for a sufficiently slow variation
of the quasicharge $q$. In this paper we show that the small-capacitance
junction model should  generally include an inductance term (see Fig.\,1b).
Such a term is proportional to the second time-derivative $\ddot{q}=\dot{I}$,
so the corresponding inductance, which we will call the "Bloch inductance,"
plays the role of the mass for quasicharge $q$. We evaluate the effect of the
Bloch inductance and show that it can be essential even for the single-band
motion of the system.

\begin{figure}[b]
\begin{center}
\includegraphics[width=3.2in]{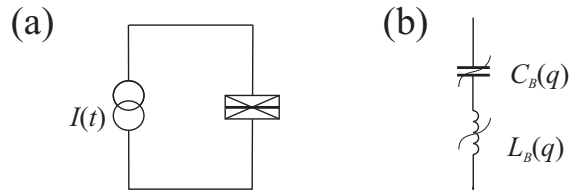}
\caption{(a) Electric diagram of the current-biased small-capacitance Josephson
junction and (b) the equivalent circuit of this junction for a small ac signal.
Connected in series the Bloch capacitance $C_B$ and Bloch inductance $L_B$ both
depend on the instant value of the quasicharge $q(t)=\int^{t}I(t') dt'$ which
is set by the current source.} \label{EqvSchm}
\end{center}
\end{figure}

The circuit consisting of a small-capacitance Josephson junction biased by the
ideal source of the classical current $I(t)$ is described by the Hamiltonian
$H=H_0+H_I$ \cite{LikZor,AZL}, where the junction's Hamiltonian and the term
due to the source are respectively equal to
\begin{equation}
\label{H_0} H_0 = \frac{Q^2}{2C}-U_J(\varphi)\quad \textrm{and}
\quad H_I=-\frac{\Phi_0}{2\pi}\varphi I(t).
\end{equation}
The operator of the charge $Q$ is conjugate to the phase operator $\varphi$ and
is equal to $Q=\frac{2e}{i}\frac{\partial}{\partial \varphi}$ in
$\varphi$-representation. The quasiparticle tunneling is neglected for the sake
of clarity. The eigenenergies $E_n(k)$ and the Bloch eigenstates $|k,n\rangle$
of the base Hamiltonial $H_0$ are periodic functions of the continuous
quasi-wavenumber $k=q/2e$ or the quasicharge $q$, with the period equal to 1
and 2$e$, respectively; the band index $n = 0, 1, 2, ...$ \cite{Box-comment}.

The operator of the phase is given in this basis by the matrix elements
\cite{LikZor,AZL,Lifshitz}
\begin{equation}
\label{phi} \varphi^{nn'}_{kk'} = i\frac{\partial}{\partial
k}\delta(k-k')\delta_{nn'} + \varphi^{nn'}_{k} \delta(k-k')(1-\delta_{nn'}).
\end{equation}
Since Hamiltonian $H$ is diagonal with respect to the variable $k$, the density
matrix elements $\rho^{nn'}_{kk'}$ take the form $\sigma_{nn'}\delta(k-k')$.
After taking trace over this variable, the equation of motion $\dot{\rho} =
(1/i \hbar)[H,\rho]$ is reduced to
\begin{equation}
\label{rho} \dot{\sigma}_{nn'} = i\omega^{nn'}_{k}\sigma_{nn'}+i
f_I\sum_{n_1=0}^{\infty}
 \left[\varphi^{nn_1}_{k}\sigma_{n_1n'} -
 \varphi^{n_1n'}_{k}
 \sigma_{nn_1}\right].
\end{equation}
Here $\omega^{nn'}_{k} = [E_{n}(k)-E_{n'}(k)]/\hbar$ is the interband angular
frequency, $f_I$ is the instant Bloch frequency $f_I= I(t)/2e$, and the
quasi-wavenumber time dependence is governed by $f_I$,
\begin{equation}
\label{k_t} k(t) = k(0)+ \int^{t}_0 f_I(\tau) d\tau.
\end{equation}

The operator of voltage, $V=(\Phi_0/2\pi)\dot{\varphi}=(i/2e)[H,\varphi]$, is
given by the matrix elements \cite{Lifshitz,Z-JETP}
\begin{equation}
\label{V_repr} V_{nn'} = \frac{1}{2e}\frac{\partial E_n}{\partial k}
\delta_{nn'}+i\frac{\Phi_0}{2\pi}\omega^{nn'}_{k}\varphi^{nn'}_k.
\end{equation}
The diagonal term $V_{00}=V(k)$ gives the voltage value in the ground state
(see the plots for a wide range of parameter $\lambda$ in Fig.\,2b of
Ref.\,\cite{Z-prl96}), while its derivative gives the reverse capacitance value
$C_B^{-1}$ Eq.\,(\ref{C_B}) shown in Fig.\,2a. The observable value of the
voltage is equal to
\begin{equation}
\label{V_obs} \langle V \rangle = \textrm{Tr}\{V\rho\} = \sum_{n,n'=0}^{\infty}
V_{nn'}\sigma_{n'n},
\end{equation}
so the non-zero off-diagonal terms of the voltage operator can also, in
principle, contribute to $\langle V \rangle$.

We solve Eq.\,(\ref{rho}) taking into account the two lowest bands, i.e. $n,n'=
0$ and 1. We also assume that the excitation to the state $n=1$ is small, so
$\sigma_{00}\lesssim 1$, while $\sigma_{11}, |\sigma_{01}|,|\sigma_{10}| \ll
1$. This assumption is adequate if both the instant Bloch frequency $f_I(t)$
and the maximum rate of its alteration $\dot{f}_I(t)$ are sufficiently small,
i.e. the dimensionless parameters $\alpha\equiv \max(f_I)/\Omega_k \ll 1$ and
$\alpha_1\equiv \max(\dot{f}_I)/\Omega_k^2 \ll 1$, where the transition
frequency is denoted as $\Omega_k \equiv \omega^{10}_{k}$. In this case, the
solution of the equations of motion
\begin{equation}
\label{r} \dot{r} \equiv \dot{\sigma}_{00}-
\dot{\sigma}_{11}=2if_I(\varphi^{01}_k \sigma_{10} - \varphi^{10}_k
\sigma_{01}),
\end{equation}
\begin{equation}
\label{sigma} \dot{\sigma}_{01} = \dot{\sigma}_{10}^*=-i\Omega_k \sigma_{01} -
if_I\varphi^{01}_k r
\end{equation}
reads
\begin{equation}
\label{r-sol} r =\sigma_{00}=1,\quad \sigma_{11}=0,
\end{equation}
\begin{eqnarray}
\label{sigma-01}
&&\sigma_{01}(t)=\sigma_{10}^*(t)=e^{-i\int_0^{t}\Omega_{k'}dt'}
 \sigma_{01}(0) \nonumber\\
 & &-ie^{-i\int_0^{t}\Omega_{k'}dt'}\int_0^{t}e^{i\int_0^{t'}\Omega_{k''} dt''}
 \varphi^{01}_{k'} f_BI(t')dt',
\end{eqnarray}
where we introduce the quasi-wavenumbers $k'=k(t')$ and $k''=k(t'')$. Keeping
the first-order contribution with respect to $\alpha$ and $\alpha_1$, we
finally arrive at
\begin{equation}
\label{sig-01-2}
\sigma_{01}=Ae^{-i\int_0^{t}\Omega_{k'}dt'}-\frac{\varphi^{01}_k}{\Omega_k}
\left(f_I(t)+i\frac{\dot{f}_I(t)}{\Omega_k}
\right).
\end{equation}
The first term describes the time-evolution (oscillations) of the density
matrix element due to Hamiltonian $H_0$. The amplitude $A$ of these
high-frequency oscillations depends on the initial conditions, i.e.
\begin{equation}
\label{A} A= \sigma_{01}(0) +
\frac{\varphi^{01}_{0}}{\Omega_0}\left(f_I(0)+i\frac{\dot{f}_I(0)}{\Omega_0}
\right),
\end{equation}
where $\varphi^{01}_{0} \equiv \varphi^{01}_{k}(t=0)$ and $\Omega_0 \equiv
\Omega_k(t=0)$. For the system initially prepared in the ground state,
$\sigma_{nn'}(0) = \delta_{n,0}\delta_{0,n'}$, and at "smooth" switching-on of
the current $I(t)$, i.e. $I(0)=0$ and $\dot{I}(0)=0$, the amplitude $A=0$. The
last term on the right-hand side of Eq.\,(\ref{sig-01-2}) is, however,
essential, yielding the dependence of $\sigma_{01}$ on $\dot{f}_I = \ddot{k}$.

The observable value of the voltage Eq.\,(\ref{V_obs}) is equal to
\begin{equation}
\label{V_obs2} \langle V \rangle =V_{00}\sigma_{00}+2\Re(V_{10}\sigma_{01}) =
V(q) + \frac{\Phi_0}{\pi}\frac{|\varphi^{01}_k|^2}{\Omega_{k}} \ddot{k}
\end{equation}
and can therefore be presented in the form
\begin{equation}
\label{V_obs3} \langle V \rangle = V(k) + L_B(k)\frac{dI}{dt}.
\end{equation}
Here we introduce the Bloch inductance
\begin{equation}
\label{L_B-def} L_B(k) = \frac{\hbar}{e^2}\frac{|\varphi^{01}_k|^2}{\Omega_{k}}
= 2\frac{E_J}{\hbar\Omega_k}|\varphi^{01}_k|^2 L_{J0},
\end{equation}
where $L_{J0}$ is the Josephson inductance of a similar classical junction
(i.e. in the absence of charging effects, $E_c \rightarrow 0$), taken at zero
argument, $L_J(\varphi=0)=\Phi_0/(2\pi I_c)$. One can see that in contrast to
the phase-dependent Josephson inductance $L_J(\varphi)$, the Bloch inductance
$L_B(k)$ is a positive periodic function of the quasi-wavenumber (quasicharge).
Physically, $L_B$ still characterizes the kinetic properties of supercurrent
which is now the operator \cite{Z-JETP} whose value depends on the system
state, i.e. on the good variable $k$, or, equivalently, a very wide wave packet
over phase $\varphi$ in $\varphi$-representation.

Let us consider the behavior of the Bloch inductance at small and large
$\lambda$. In these limiting cases, the analytical expressions for the matrix
elements $\varphi^{01}_k$ can be taken from Ref.\,\cite{LikZor}. For $\lambda
\ll 1$, inductance $L_B$ increases resonantly in the vicinity of the degeneracy
points, i.e.
\begin{equation}
\label{q=e}   L_B =  L_B^{\textrm{max}}/(1+\xi^2)^2, \quad
|\xi|\lesssim 1
\end{equation}
at $k= (1+\lambda\xi)/2 \:(\textrm{mod}\:1)$. The peak value is equal to
$L_B^{\textrm{max}}= 32 \lambda^{-2} L_{J0} \gg L_{J0}$. Outside the resonance
region, $L_B$ drops dramatically approaching at $k \approx 0
\:(\textrm{mod}\:1)$ the lowest level of $L_B^{\textrm{min}} = 0.5 \lambda^3
L_{J0} \ll L_{J0}$.

\begin{figure}[t]
\begin{center}
\includegraphics[width = 3.4in]{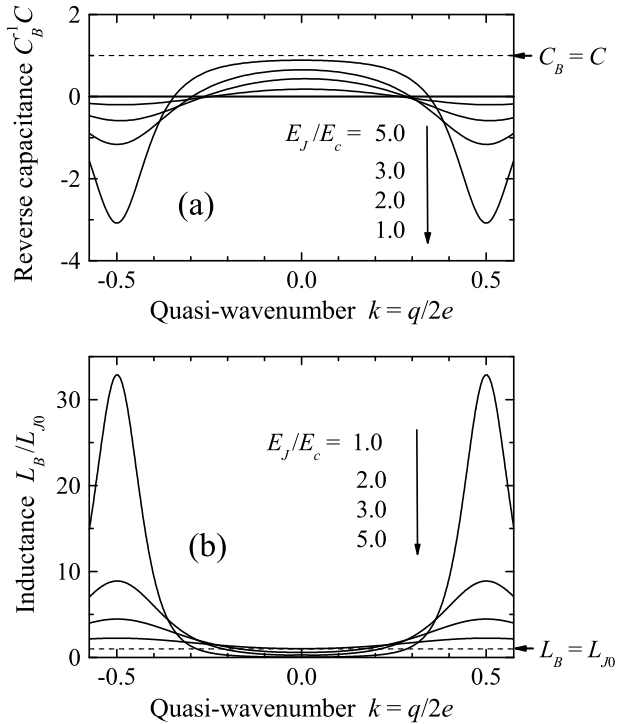}
\caption{Normalized reverse Bloch capacitance (a) and Bloch inductance (b) in
the zero band (ground state) versus quasicharge for several values of the
energy ratio $\lambda=E_J/E_c$.} \label{ResCurve}
\end{center}
\end{figure}

At large Josephson couplings, $\lambda \gg 1$, both the interband frequency
$\Omega_{k} \approx \omega_p = (8E_JE_c)^{1/2}/\hbar
=(8/\lambda)^{1/2}E_J/\hbar$ ($\omega_p$ is the Josephson plasma frequency) and
the matrix elements $|\varphi^{01}_k| \approx |\varphi^{01}_{\textrm{osc}}|=
(2/\lambda)^{1/4}$ (equal to those of the equivalent harmonic oscillator) do
not depend on $k$. The Bloch inductance takes on almost the constant value,
$L_B \approx L_{J0}$. Taking into account the fact that corrections to
$\varphi^{01}_{\textrm{osc}}$ are exponentially small ($\propto \exp \left[
-(8\lambda)^{1/2} \right]$) and accounting only for corrections to the
oscillator's energy eigenvalues due to anharmonicity of the Josephson
potential, the expression for $L_B$ reads
\begin{equation}
\label{L_B-ll} L_B(k) = \left[ 1 + (2\lambda)^{-1/2} \right]
L_{J0}.
\end{equation}
For several intermediate values of parameter $\lambda$, the inductance $L_B$ is
presented by the plots in Fig.\,2b, obtained by numerical methods.

If the dc part of the quasi-wavenumber $k =q/2e$ is fixed and the small ac part
is driven by current $I(t)=I_1 e^{i\omega t}$, the junction voltage according
to Eq.\,(\ref{V_obs3}) is equal to $V= V(q)+ Z(\omega)I_1 e^{i\omega t}$. The
electrical impedance of the junction $Z(\omega)=i[\omega L_B - (\omega
C_B)^{-1}]$ is purely reactive (see Fig.\,1b). So, at a sufficiently high
irradiation frequency $\omega$, the inductive term can be appreciable.

At small Josephson couplings, $\lambda\ll 1$, the maximum effect of $L_B$ can
be expected to be near the degeneracy points Eq.\,(\ref{q=e}). However, even in
these points the inductive resistance $\omega L_B$ approaches the capacitive
resistance $-(\omega C_B)^{-1}$ (the positive value due to the negative value
of $C_B = - \lambda C/4$) only at $\omega = \Omega_k=E_J/\hbar$, i.e. at the
frequency of the interband transition. At such a high frequency our approach is
not applicable. At lower frequencies, the effect of $L_B$ is rather small and
can be presented as a frequency-dependent correction to the reverse Bloch
capacitance,
\begin{equation}
\label{C_B-corr}C_B^{-1}=-\frac{4}{\lambda C} \,\rightarrow\,
-\frac{4}{\lambda C}
\left(1+\frac{\omega^2}{\Omega_k^{2}}\right), \quad \omega \ll
\Omega_k.
\end{equation}
Such renormalization of the Bloch capacitance can, in principle, be detected in
experiment on the high-frequency Bloch-capacitance-based readout of the Cooper
pair box qubit \cite{Duty}.

For $\lambda \gg 1$, as follows from the harmonic shape of the eigenenergy
$E_0(q)$ (see Eqs.\,(A3)-(A4) of Ref.\,\cite{LikZor}), the reverse Bloch
capacitance is small,
\begin{equation}
\label{C_B-ll} C_B^{-1}(k) = C_{B0}^{-1}\cos(2\pi k), \quad C_{B0}^{-1}=b \,
B(\lambda) \,C^{-1},
\end{equation}
\begin{equation}
\label{L_B-ll2}B(\lambda)=\lambda^{3/4}\exp \left[
-(8\lambda)^{1/2} \right]\ll 1,
\end{equation}
where the numeric factor $b=\pi^{3/2}2^{11/4} \approx 37$. The maximum
(resonance) frequency above which the inductor dominates, is
\begin{equation}
\label{omega-res} \omega_0 =[L_B(0)C_{B0}]^{-1/2} = (bB)^{1/2}\omega_p \ll
\Omega_k.
\end{equation}
Therefore, at least in the range of the frequencies $\omega_0 \lesssim \omega
\ll \Omega_k$, the effect of inductance $L_B$ should be significant.

The question arises whether the role of the Bloch inductance is essential in
the dynamics of an autonomous Josephson junction. For evaluating this effect we
introduce a small dissipation by adding to our model Eq.\,(\ref{H_0}) a
high-ohmic resistor, $R \gg R_Q\equiv h/4e^2 \approx 6.45$\,k$\Omega$,
connected either parallel to the current source in Fig.\,1 \cite{LikZor,AZL}
or, equivalently, in series with a voltage source $V_0$, as shown in the inset
in Fig.\,3. In this case the system is described by a wave packet over $k$ (see
Eqs.\,(71) and (75) of Ref.\,\cite{LikZor}). At low temperature, $k_BT \ll
\min\{ \hbar \Omega_k,[E_0(e)-E_0(0)]\}$, this packet with the center $k$, is
sufficiently narrow, $\Delta k \ll 1$. Corresponding interband damping terms in
the equation of motion Eq.\,(\ref{sigma}) lead to an exponential decay of the
fast-oscillating off-diagonal term $\propto A$ in Eq.\,(\ref{sig-01-2}).
Finally, the equation of motion for $q$ takes the form
\begin{equation}
\label{model-eq} L_B(q)\ddot{q}+R\dot{q}+V(q)=V_0,
\end{equation}
or, in dimensionless units,
\begin{equation}
\label{model-dimless} \ell(\theta) \beta_L \frac{d^2
\theta}{d\tau^2}+\frac{d\theta}{d\tau}+g(\theta)=v.
\end{equation}
Here we introduced $\theta = 2\pi k$, $\tau =  \omega_c t$, $\omega_c=V_c/eR$,
$ v=V_0/V_c$, $V_c = \max[V(q)]$ and the unity-amplitude periodic function
$g(\theta)=V(e\theta/\pi)/V_c$. Parameter $\beta_L = L_{J0}V_c/eR^2 $ and the
normalized Bloch inductance is $\ell(\theta) = L_B(e\theta/\pi)/L_{J0}$. For
large $\lambda$, the values of these parameters are $V_c\approx
\pi^{-1}eC_{B0}^{-1}$, $\omega_c \approx (RC_{B0})^{-1}$, $g(\theta) \approx
\sin \theta$, $\ell(\theta)\approx 1$ and
\begin{equation}
\label{beta} \beta_L \approx \frac{\omega_c^2}{\omega_0^2}= \frac{L_{J0}}{R^2
C_{B0}}= \frac{2bB(\lambda)}{\pi^2 \lambda} \left(\frac{R_Q}{R}\right)^2 \ll 1,
\end{equation}
and Eq.\,(\ref{model-dimless}) takes the form of the equation, describing a
driven, damped pendulum or a resistively shunted Josephson junction \cite{RSJ}.
Finite values of corresponding McCumber-Stewart parameter, $\beta_L \gtrsim 1$,
characterizing the effect of the inductive term on the dc $I$-$V$ curve (IVC),
can therefore not be achieved.

However, the situation changes dramatically in the case of an array of $N$
junctions connected in series in the same network (see the central part of the
inset in Fig.\,3). The individual junctions of this array are decoupled and
described by Eqs.\,(\ref{model-eq})-(\ref{beta}) with replacement $L_B(q)
\rightarrow N L_B(q)$, $V(q) \rightarrow N V(q)$, $V_c \rightarrow N V_c$ and
$\beta_L \rightarrow \beta_L^* = N^2\beta_L$. Note that the maximum resonance
frequency $\omega_0$ Eq.\,(\ref{model-dimless}) does not change. At
sufficiently large $N\gg 1$, the effective McCumber-Stewart parameter
$\beta_L^*$ can be made of the order of 1, even for unavoidably large values of
$R$. As a result, the IVC of such an array can exhibit characteristic
hysteresis (see Fig.\,3).

\begin{figure}[t]
\begin{center}
\includegraphics[width = 3.4in]{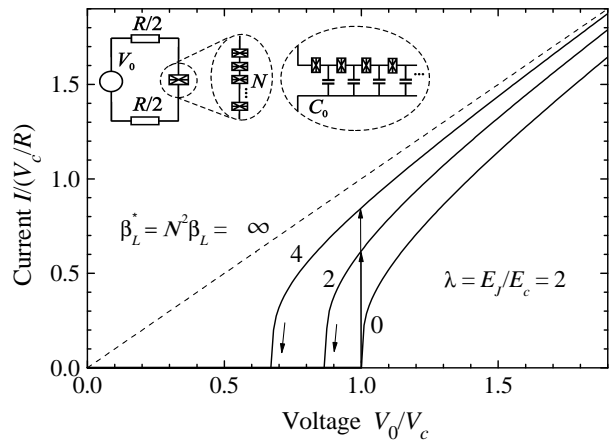}
\caption{$I$-$V$ characteristics obtained in the model including the
quasicharge-dependent Bloch inductance
Eqs.\,(\ref{model-eq})-(\ref{model-dimless}) for several values of parameter
$\beta_L^*$. The curves corresponding to the finite values of $\beta_L^*$ show
appreciable hysteresis. The inset shows possible realizations of large values
of $\beta_L^*$.} \label{IVC}
\end{center}
\end{figure}

In experiment, due to the small stray capacitance $C_0$ of each island of the
array to ground (see the rightmost part in the inset in Fig.\,3), the number of
decoupled junctions $N_0$ is limited by the size of the Cooper-pair soliton,
i.e. $N_0 = (C_{B0}/C_0)^{1/2}$ \cite{q-SG,HavDel}. At $\lambda \gg 1$, this
size increases dramatically, so that it can result in hysteretic IVC. Similar
effect has been observed repeatedly by KTH group in long arrays of small Al
junctions (see, e.g., \cite{Agren}), although these arrays were voltage-biased,
so the quasiparticle tunnelling played an important role in those samples.

The concept of the Bloch inductance allows the sine-Gordon equation, describing
static Cooper-pair 2$e$-solitons in the small-capacitance junction array
\cite{q-SG}, to be generalized to the non-stationary case. Each junction
included in the elementary cell of the array is presented as shown in Fig.\,1.
Due to the emerging inertial term $L_B\ddot{q}$, the corresponding sine-Gordon
equation for the quasicharge describes a persistent motion of the
2$e$-solitons.

In conclusion, we introduced the Bloch inductance and described its effect on
the quasicharge dynamics in circuits with small-capacitance Josephson
junctions. The effect should manifest itself by an effective renormalization of
the Bloch capacitance in the vicinity of the degeneracy points of the Cooper
pair boxes and by a hysteretic behavior of the IVCs of long one-dimensional
arrays. The Bloch inductance of the Josephson junction included in the variable
electrostatic transformer \cite{AvBrud} may at $\lambda \gg 1$ result in an
inter-qubit coupling of resonator-mediated type. Finally, similar to the
Josephson inductance the Bloch inductance may be applied for constructing the
dispersive QND-type readout of Josephson qubits \cite{Z-Phys-C}.

I wish to thank D.\,B.~Haviland and S.\,V.~Lotkhov for stimulating discussions
and G.~Sch\"{o}n and D.~Esteve for comments. This work was partially supported
by the EU through the SQUBIT-2 and RSFQubit projects.


\begin{references}
\bibitem{Box-comment} After replacing the current source $I$ (in fact, the
charge source) by a voltage source $V_0$ with small capacitance $C_g$ connected
in series, this model describes the Cooper pair box. In this case, the
quasicharge is equal to the polarization charge on the capacitive gate, i.e.
$q=C_g V_0$.
\end{references}
\end{document}